\documentclass[11pt,twoside]{article}
\usepackage{mattstyle}  
\usepackage[utf8]{inputenc}

\newcommand{\possessivecite}[1]{\citeauthor{#1}'s (\citeyear{#1})}

\DeclareMathOperator{\arsinh}{arsinh}

\graphicspath{ {figures/} }

\usepackage[flushleft]{threeparttable} 

\makeatletter
\def\thanks#1{\protected@xdef\@thanks{\@thanks
        \protect\footnotetext{#1}}}
\makeatother

\usepackage[section]{placeins}

\usepackage[super]{nth}

\usepackage{pdflscape}

\begin{document}


\begin{spacing}{1}
\begin{titlepage}

\author{Joshua E. Blumenstock\\{\small UC Berkeley}
\and Matthew Olckers\\
{\small UNSW Sydney}\thanks{We thank Milo Bianchi, Jonathan Guryan, Kai Barron, and Xiaojian Zhao for helpful and detailed feedback. This project started while Matthew was a Ph.D. student at the Paris School of Economics and a visiting student researcher at UC Berkeley. We are grateful to researchers from both institutions for helpful comments. In particular, we thank Francis Bloch, Margherita Comola, Fabrice Etilé, Simon Gleyze, Sylvie Lambert, Liam-Wren Lewis, Karen Macours and David Margolis. \newline \newline Blumenstock: jblumenstock@berkeley.edu; Olckers: m.olckers@unsw.edu.au}}

\title{Gamblers Learn from Experience}
\runningheads{Blumenstock \& Olckers}{Gamblers Learn from Experience}

\date{\small \today }
\maketitle
\thispagestyle{empty}

\vspace{-0.3in}

\begin{abstract}

Mobile phone-based sports betting has exploded in popularity in many African countries. Commentators worry that low-ability gamblers will not learn from experience, and may rely on debt to gamble. Using data on financial transactions for over 50\,000 Kenyan smartphone users, we find that gamblers do learn from experience. Gamblers are less likely to bet following poor results and more likely to bet following good results. The reaction to positive and negative feedback is of equal magnitude and is consistent with a model of Bayesian updating. Using an instrumental variables strategy, we find no evidence that increased gambling leads to increased debt.
\end{abstract}

\end{titlepage}
\end{spacing}



\section{Introduction}

Enabled by mobile phone technology, sports betting has exploded in popularity in many African countries, such as Kenya. For instance, the term ``SportPesa'', the name of Kenya's largest betting platform, was the most popular search term on Google in 2018.%
\footnote{Of the top five search queries, three were related to sports betting (``sportpesa", ``livescore",  ``betpawa"). The other two were ``kenya" and ``news.'' Facebook was sixth, with only a quarter of the search volume as SportPesa. See \href{https://trends.google.com/trends/explore?date=2018-01-01\%202018-12-31&geo=KE}{trends.google.com}} %
A 2017 GeoPoll survey of 3\,879 youth in sub-Saharan Africa found that 76\% of Kenyan youth had used their mobile phone for gambling \citep{geopoll2017}. These betting apps are designed to work on even the most basic mobile phones, thus creating new opportunities to gamble for millions of low-income individuals.

Mobile phone-based sports betting has spread from Kenya, an early adopter of mobile money, throughout Africa and other low and middle-income countries. Sports betting is also popular in wealthy nations, with the global sports betting market expected to grow by 140 billion dollars by 2024 \citep{technavio_global_2020}. In the United States alone, where sports betting was only recently legalized, an estimated 8 billion dollar market is projected by 2024 \citep{hancock_bet_2019}.

The rapid growth of sports betting has stoked concerns that sports betting could drive people into financial ruin---especially when credit is easily available. Commenting on the recent legalization of sports betting in the United States, Victor Matheson, an expert on the economics of sports, raises concerns about whether gamblers can learn from experience.\footnote{See \href{https://gen.medium.com/the-economics-of-sports-gambling-4ee5fa7d7b9e}{Freakonomics podcast}, episode 388.}%

\begin{quotation}
\noindent \textit{``Think of all those sports fans who say, `You know, I would never buy a lottery ticket. That’s just luck. But I know everything about sports. I should be able to win this.' And guess what? You can’t beat the casino. These amateurs who think they’re experts don’t stand a chance, but do stand a chance of really getting sucked in. And the question is how quickly can they extricate themselves and realize that ‘Yeah, I’m actually not any good at this.' "}
\end{quotation}

This paper shows that gamblers learn from experience. Gamblers are less likely to bet following negative feedback on their gambling ability and more likely to bet following positive feedback. The magnitude of the reaction is similar for positive and negative feedback.

We demonstrate this result using a rich dataset that captures all mobile money transactions, including detailed information on sports bets and consumer loans, for thousands of Kenyan mobile phone owners. We study how gamblers respond to experience by focusing our analysis on one extremely popular type of bet, the SportPesa `jackpot bet',  where the gambler simultaneously makes predictions on 17 soccer matches. The gambler only wins money if he or she predicts 12 or more outcomes correctly. By focusing on how the gambler reacts to his or her performance below the 12-match threshold, we separate the signal of correct predictions from the income effect of winning a bet.

Our first set of results show that gamblers increase betting expenditure in weeks after making successful predictions and reduce betting expenditure after making incorrect predictions. We find gamblers respond with equal magnitude to positive and negative feedback. Gamblers are 2.8 percentage points more likely to place a bet when they predict a relatively high share of matches correctly and 2.9 percentage points less likely to place a bet when they predict a relatively low share of matches correctly. The symmetric response persists for at least three weeks.

The symmetry in response to positive and negative feedback stands in contrast to conventional wisdom. To indicate how people expect gamblers to behave, we surveyed 32 academics and business professionals before writing this paper. The majority (18 respondents) predicted gamblers would be more responsive to positive than to negative feedback. Only 4 respondents predicted the correct result---that gamblers respond equally to positive and negative feedback.

Our second set of results investigate the effects of increased betting on other financial activities of the gambler. We adopt an instrumental variables strategy where we use random variation in gambling outcomes from prior weeks to instrument for the gambler's current betting expenditure. Our identifying assumption is that, conditional on the number of bets placed and individual fixed effects (which control for gambling ability and other unobserved individual characteristics), the number of correct predictions in a given week is random.

Analyzing financial transactions in the week of a small positive shock to gambling expenditures, we find that gamblers more actively make deposits and withdrawals from their savings account, but observe no significant net effect on savings. Analyzing loan applications and repayments, we find that the shocks to gambling expenditures do not cause gamblers to apply for consumer credit. These results stand in contrast to a popular narrative that borrowers are gambling on credit \citep{economist2018}.
\vspace{.2cm}

These results contribute to an active debate about the role and regulation of sports betting in developing economies. The negative expected returns from gambling have motivated policymakers to discourage gambling.%
\footnote{In Kenya, the government has introduced taxes on revenue and winnings and the gambling regulator has introduced bans on outdoor advertising and celebrity endorsements. In Tanzania, religious leaders have pushed for sports betting to be banned altogether. The Ugandan government has suspended all new gambling licenses and pledged not to renew existing licenses.} %
Recent field experiments, which test whether gambling demand can be reduced by better informing gamblers of the likelihood of winning, have shown mixed results \citep{zenker2018,abel2020}.%
\footnote{In South Africa, researchers asked participants to roll sixes with dice to demonstrate the tiny probability of winning the national lottery, and found  that those who took longer to get a six decreased demand for lottery tickets \citep{abel2020}.  The experiment in Thailand demonstrated the probability of winning the national lottery with a poster containing one million dots representing the number of tickets sold and several pins representing number of winning tickets per prize category \citep{zenker2018}. The demonstration lead to improved knowledge but no change in the willingness to pay for lottery tickets. One way to reconcile these results would be if an individual's belief in their own probability of winning differed from their  belief about the population average. \possessivecite{abel2020} experiment, which found changes in behavior, focused on the individual whereas \possessivecite{zenker2018} experiment, which found no change in behavior, focused on the population average.} %

The results complement studies of people's motivation to gamble. A recent field experiment in Uganda suggests that sports bettors gamble to raise funds for large lump sum purchases \citep{herskowitz2019}. Sports betting may serve as a substitute for credit---again a contrast to the popular narrative that gambling drives people into debt.

These findings also relate to studies of investor behavior and stock market speculation. Sports betting shares a similar structure to stock market speculation \citep{sauer1998wagering,levitt2004}. And like sports betting, the large number of small investors who make low or even negative returns has ignited much debate on whether these investors learn rationally from experience or with a bias \citep{barber2020}. Our setting allows us to separate feedback on ability from income effects. We find support for a rational model of learning \citep{mahani2007,linnainmaa2011}.

More broadly, our results provide insight into how people respond to signals of their ability.  Lab experiments find that gamblers tend to accept correct predictions but explain away incorrect predictions \citep{gilovich1983biased}.%
\footnote{People attribute success to their own skill and explain away failure, even when the task has no element of skill---such as flipping a coin \citep{langer1975heads}.} %
Recent literature has found that people react more to positive than to negative feedback \citep{eil2011good,sharot2011,mobius2014,zimmerman2020} and tend to forget negative feedback \citep{chew2019motivated, zimmerman2020,godker2019,huffman2020}. Some studies do find people react more to negative feedback \citep{ertac2011,coutts2019}, especially concerning an investment decision \citep{kuhnen2015}. We find that gamblers react symmetrically to good and bad results in the previous week, which supports a model of Bayesian updating.%
\footnote{Symmetric updating does not rule out biased priors \citep{grossman2012unlucky,buser2018}. In our data, we cannot observe the gambler's prior on his or her betting ability.} %
 A recent experiment focusing only on financial decisions also finds support for Bayesian updating \citep{barron2020}. Perhaps ability in financial contexts, such as sports betting, may not be important for self-image \citep{gotthard2017}. Therefore, gamblers do not need to discount negative feedback about their gambling ability to retain a positive self-image.\footnote{\citet{benjamin2019errors} emphasizes that the evidence on biased updating is ``confusing" and there is no unified explanation for the range of findings.}

In summary, our paper makes three contributions. First, we use a novel empirical setting to provide evidence that gamblers do learn from experience. The structure of the jackpot bets allows us to separate feedback on ability from income effects. Second, in contrast to a growing literature on biased updating, we find that gamblers display responses of equal magnitude to positive and negative feedback. Third, we use the relationship between past betting results and current betting expenditure to provide suggestive evidence of how increases in betting expenditure affect the use of savings and credit.

\section{Model of Gambling Behavior} \label{section:model}

We provide a model of gambling behavior to study how gamblers learn from experience.

A gambler can predict the outcome of a match with probability $\theta \in [0,1]$, which represents the gambler's ability.
Sports betting differs from other gambles such as lotteries in that $\theta$ can differ between gamblers. In sports betting, some gamblers may have more skill in determining the outcome of a match, whereas, in a lottery, the probability of winning is identical for all gamblers.

The gambler does not know his ability. He has a prior belief $\Pr[\theta] \sim \operatorname{Beta}(s_w, u_w)$. Using the beta distribution to specify the gambler's prior allows for an intuitive interpretation. Suppose, before starting to gamble, the gambler watches $m_w$ matches to test how many he can predict correctly. His prior belief of $\theta$ has the distribution $\operatorname{Beta}(s_w, u_w)$ if the gambler predicts the outcome of $s_w$ matches successfully and $u_w$ matches unsuccessfully.\footnote{We assume a Haldane prior for gamblers who have never watched any soccer matches before.}

Each week, a gambling operator offers a bet on the outcome of a sports match. The gambler chooses whether to bet as a function of his expected ability, $\mathrm{E}[\theta]= \frac{s_w}{m_w}$ (where $m_w=s_w+u_w$).%
\footnote{We could also include other moments of the gambler's belief of his ability in the model. For example, the variance of his ability is $\mathrm{Var}[\theta] = \frac{(s_w+s_b)(u_w+u+b)}{(m_w+m_b)^2(m_w+m_b + 1)}$. This variance approaches zero as the number of matches the gambler watches, $m_w$, and bets on, $m_b$, increases.} %
The gambler only bets if %
\begin{align*}
\mathrm{E}[\theta] \geq c_t \ ,
\end{align*} %
where $c_t \in [0,1]$ is a cutoff in week $t$. The cutoff $c_t$ captures factors other than the gambler's belief of his ability which impacts his decision to bet. This cutoff may differ from week to week for a variety of reasons, such as the gambler's budget or his amount of free time. The cutoff may be especially low if the gambler derives utility from the act of gambling \citep{conlisk1993}. In the extreme, if $c_t=0$, the gambler will always bet regardless of his expectation of his betting ability.

Once the gambler starts betting, he uses Bayes Rule to update his belief of his betting ability $\theta$. Let $s_b$ be the number of successful bets and  $u_b$ the number of unsuccessful bets in $m_b$ matches. The number of successful bets $s_b$ is generated from a binomial distribution with $m_b$ trials and a success probability of $\theta$, the gambler's ability. Since the beta distribution is a conjugate of the binomial distribution, the posterior belief also has a beta distribution. \begin{align*}
    \Pr[\theta|s_b] &\sim \operatorname{Beta}(s_w + s_b,u_w + u_b) \\
    \mathrm{E}[\theta|s_b] &= \dfrac{s_w + s_b}{m_w+m_b}
\end{align*}
This relationship between the posterior and prior beliefs has an intuitive interpretation. The prior simply adds $s_w$ successful bets and $u_w$ unsuccessful bets to the gambler's record of $s_b$ successful bets and $u_b$ unsuccessful bets.

If the gambler fails to learn from experience, by ignoring past results or only remembering successful bets, he may never learn his ability and become trapped by his belief that he will eventually win big. Our unique data allows us to test how gamblers react to feedback. We find that gamblers do learn from experience and react in a similar way to the Bayesian gambler in our model.

\section{Context and Data: Mobile Money and Sports Betting}\label{section:context}

In 2007, the Kenyan telecom company Safaricom launched M-Pesa, a mobile phone-based financial platform that allows users to conduct basic financial transactions over the mobile phone network. Since 2007, mobile money has proliferated in the developing world and today there are more than a billion registered mobile money accounts across 290 mobile money services in at least 95 countries \citep{GSMA2019}. Currently, mobile money accounts are more common than bank accounts in most African nations.

The introduction of mobile money has been associated with important welfare effects. In Kenya, mobile money has been linked to improved risk-coping \citep{jack2014} and is estimated to have lifted as many as 194 000 Kenyans out of poverty \citep{suri2016}. Businesses also benefit from mobile phone adoption as payments for goods and services can be collected with mobile money. Consequently, many in the policy and aid communities view mobile financial services as key to improving financial inclusion among the poor \citep{GSMA2019,lauer2015}.

The widespread adoption of mobile money has been accompanied by the rapid spread of mobile phone-based gambling. Gambling operators have leveraged the M-Pesa mobile money network to collect bets on soccer matches and other sports. Since the transaction cost of collecting bets and disbursing winnings has been lowered by mobile phone technology, the companies offer bets for as little as 1 KSH (0.01 USD).

Many Kenyans have a strong interest in sport, especially soccer. Gambling operators have tapped into this interest and sports betting has been widely adopted.%
\footnote{A survey of $1\,130$ Kenyan youth between the ages of 17 and 35 conducted by GeoPoll in 2017 found that 60 percent of respondents had placed bets on football matches in the past and a further 16 percent had tried other forms of betting \citep{geopoll2017}.} %
In just five years of operation, SportPesa, the leading sports betting operator, has gathered enough revenue to sponsor two English Premiership soccer teams, deals estimated at 4 and 12 million dollars per year.%
\footnote{A \href{https://nation.africa/kenya/news/shocking-details-of-sh30bn-a-month-bets-309350}{leaked spreadsheet} from Kenya's betting regulator revealed that wagers totaled 300 million USD in May 2019 and SportPesa is estimated to hold two-thirds of this market.}

We use a dataset that contains basic metadata on all mobile money transactions conducted by an anonymized sample of Kenyan smartphone owners. These data were collected by a smartphone app, installed by 93\,565  Kenyans between April and June 2016 and in July 2017. No personally identifiable information was used in this study.

We use a subsample of 58\,215 individuals for whom we observe a payment to a gambling company. The large number of gamblers in our sample matches survey evidence on gambling incidence in Kenyan youth \citep{geopoll2017}. Among the subsample of gamblers, 77 percent are men and 75 percent are 35 years of age or younger. Appendix~\ref{appendix:descriptive_stats} contains additional descriptive statistics.

This sample provides insight into the behavior of working-age adults in African urban areas---where the growth of sports betting has been most rapid. Although sports betting is popular worldwide, these findings may not generalize to contexts where sports betting systems and behaviors are markedly different from the Kenyan context.

Our analysis focuses on transactions between individuals and private companies. For each transaction, we observe the value of the transaction and the name of the company involved. Since sports betting in Kenya operates almost exclusively via mobile money, we can observe the deposits and withdrawals the gambler makes.

We also observe the details of transactions with the betting company, SportPesa. In a typical transaction, we observe the type of bet placed, the picks the gambler made, and the amount wagered. For example, a transaction may show \texttt{You've placed Jackpot BetID abc123 2\#1\#2\#x\#2\#1\#1\#1\#2\#x\#1\#1\#x\#1\#2\#1\#2 for KSH 100. S-Pesa available balance KSH 100. Games resulted on full time.}. The string of characters and hashes represent the gambler's picks on each of the 17 games in the jackpot. The \texttt{1} is for home team wins, the \texttt{2} for away team wins, and \texttt{x} for a draw.


\section{Measuring Gambling Ability with Jackpot Performance}\label{section:ability}

Our model of gambling assumes that there is heterogeneity in gambling ability, such that high ability gamblers have a greater probability of predicting the outcome of sports matches correctly than low ability gamblers. To support this assumption, we show that certain gamblers consistently predict more matches correctly than others.

\subsection{Jackpots as a Standardized Measure of Gambling Ability}\label{section:jackpots}

One empirical challenge is to define a standardized measure of betting ability. Sports betting companies offer a range of different bets. For example, gamblers can bet on single matches or they can chain matches together to increase the payout (and decrease the probability of winning). Also, individual matches differ in odds. It is easier to predict the outcome of a match with a clear favorite than to predict the outcome of a match with similar strength teams. We address this challenge by studying the number of correct predictions on the weekly jackpots offered by SportPesa, Kenya's leading sports betting operator.

Each week, SportPesa selects 13 matches for a midweek jackpot and 17 matches for a weekend jackpot (called the MegaJackpot). The jackpot is awarded if the outcomes of all the pre-selected matches are predicted correctly. Bonuses are awarded for 10 or more correct predictions on the midweek jackpot and 12 or more correct predictions on the weekend jackpot. The amount of the bonus increases exponentially with the number of correct predictions. For example, the bonuses for the weekend jackpot on 31 March 2019 were approximately USD \$460, \$2\,190, \$8\,140 and \$64\,300 for 12, 13, 14 and 15 correct predictions.

The jackpots provide a standardized measure of betting ability, both between gamblers and across time. Since SportPesa selects the matches each week we need not worry about how the gambler picks the matches. The types of matches SportPesa selects each week are relatively obscure, and generally have no clear favorite, such that the odds are balanced across the three possible outcomes of home team wins, away team wins, or a draw. As an example of a typical jackpot bet, Table \ref{table:example_jackpot} shows the weekend jackpot matches listed for the weekend of 5 June 2021.

\begin{table}[tbp] \centering
\caption{Example of a SportPesa jackpot}
\small
\begin{threeparttable}
\begin{tabular}{l c c c c}
\hline
	&	\multicolumn{3}{c}{\textbf{Odds}}		&		\\
\textbf{Match (Home vs Away)} & \textbf{Home win} & \textbf{Draw} & \textbf{Away win} & Outcome \\ \hline
Black Leopards vs Bloemfontein Celtic	&	2.87	&	2.97	&	2.63	&	Draw	\\
Maritzburg United vs Amazulu FC	&	2.58	&	2.93	&	2.98	&	Draw	\\
TS Galaxy FC vs Kaizer Chiefs	&	3.25	&	2.8	&	2.5	&	Away Win	\\
GKS Belchatow vs GKS Jastrzebie Zdroj	&	2.59	&	3.34	&	2.59	&	Away Win	\\
Orgryte IS vs IFK Varnamo	&	2.65	&	3.4	&	2.6	&	Draw	\\
Plaza Colonia CD vs Nacional (URU)	&	2.72	&	3.46	&	2.42	&	Away Win	\\
IA Sud America vs CS Cerrito	&	2.62	&	3.06	&	2.77	&	Draw	\\
Zwiegen Kanazawa vs Omiya Adija	&	2.54	&	3.13	&	2.88	&	Home Win	\\
JEF United Chiba vs Montedio Yamagata	&	2.74	&	3.09	&	2.69	&	Draw	\\
OKS Odra Opole vs GKS Tychy	&	3	&	3.25	&	2.4	&	Away Win	\\
Zaglebie Sos. vs Gornik Leczna	&	2.48	&	3.45	&	2.7	&	Away Win	\\
Falkenbergs vs Vasteras SK	&	2.9	&	3.35	&	2.41	&	Home Win	\\
RoPs Rovaniemi vs EIF Ekenas	&	2.46	&	3.17	&	2.87	&	Draw	\\
Pogon Sieflce vs LKP Motor Lublin	&	2.55	&	3.18	&	2.51	&	Draw	\\
PK-35 vs TPS Turku	&	2.87	&	3.08	&	2.63	&	Draw	\\
America Mineiro MG vs Corinthians	&	2.78	&	3.08	&	2.63	&	Away Win	\\
Germany U21 vs Portugal U21	&	2.55	&	3.21	&	2.8	&	Home Win	\\
\hline
\end{tabular}
\begin{tablenotes}[flushleft]
\footnotesize{\item \emph{Notes:} Matches for the weekend jackpot on the weekend of 5 June 2021, sourced from \href{https://twitter.com/SportPesa_Care/status/1400842205728751619}{SportPesa's Twitter account}. A gambler must predict one of three outcomes for each match: home win, away win or draw. The gambler wins money for 12 or more correct predictions on the 17 matches. We have listed the odds for each outcome, which represents the payout a gambler would receive if he bet on a single match. The higher the odds, the less likely the outcome. The inverse odds of each match do not sum to one. The difference is the betting company's margin. We have also listed the outcome of each match, which the gambler only observes after he places a bet.}
\end{tablenotes}
\end{threeparttable}
\label{table:example_jackpot}
\end{table}

The jackpots are extremely popular. In our data, we observe midweek or weekend jackpot bets placed by over 17\,000 individuals, which represents 30 percent of all individuals for whom we observe some type of betting transaction.

The popularity of the jackpot may stem from the high payouts in comparison to alternative betting options. Note that gamblers could construct an equivalent bet to the jackpot by chaining together the same matches as the jackpot. The payout from this ``multibet" is equal to the product of the odds of the predicted outcomes. Thus, if a gambler constructed a multibet for the matches listed in Table \ref{table:example_jackpot}, he would win KSH 42 million. However, the same bet through the jackpot has a payout of KSH 129 million---a prize three times larger. In addition, if some of his predictions were unsuccessful, he may still win a bonus.

\subsection{Estimates of Gambling Ability}

We directly estimate the gambler's ability $\theta$ and  find significant differences in gambling ability between gamblers. Each jackpot match provides a standardized type of bet to compare gamblers and estimate their ability. As described in Section~\ref{section:model}, the expectation of ability, $\theta$, for $s$ correct predictions is $m$ matches is $\mathrm{E}[\theta|s,m] = \dfrac{s}{m}$.  Provided $s$ is generated from a binomial distribution with a $\theta$ probability of success, $\mathrm{E}[\theta|s,m]$ follows a beta distribution. We can derive Clopper-Pearson confidence intervals around $\mathrm{E}[\theta|s,m]$ from the beta distribution.

The range of the confidence interval for a particular gambler is a function of the number of matches we observe for that gambler. We focus initially on a smaller group of gamblers for which we observe a large number of matches.%
\footnote{In Appendix~\ref{sec:dynamics}, we use the full sample of gamblers to document correlations in each gambler's betting results over time. We find that gamblers who predict a high number of matches correctly in the current week are more likely to predict a high number of matches correctly in the following week.} In this sample, the smallest confidence interval is 0.06, a gambler for whom we observe 1\,129 match predictions.

\begin{figure}[htb]
    \centering
    \caption{Estimates of betting ability for frequent gamblers \label{fig:ability}}    \includegraphics[width=\textwidth]{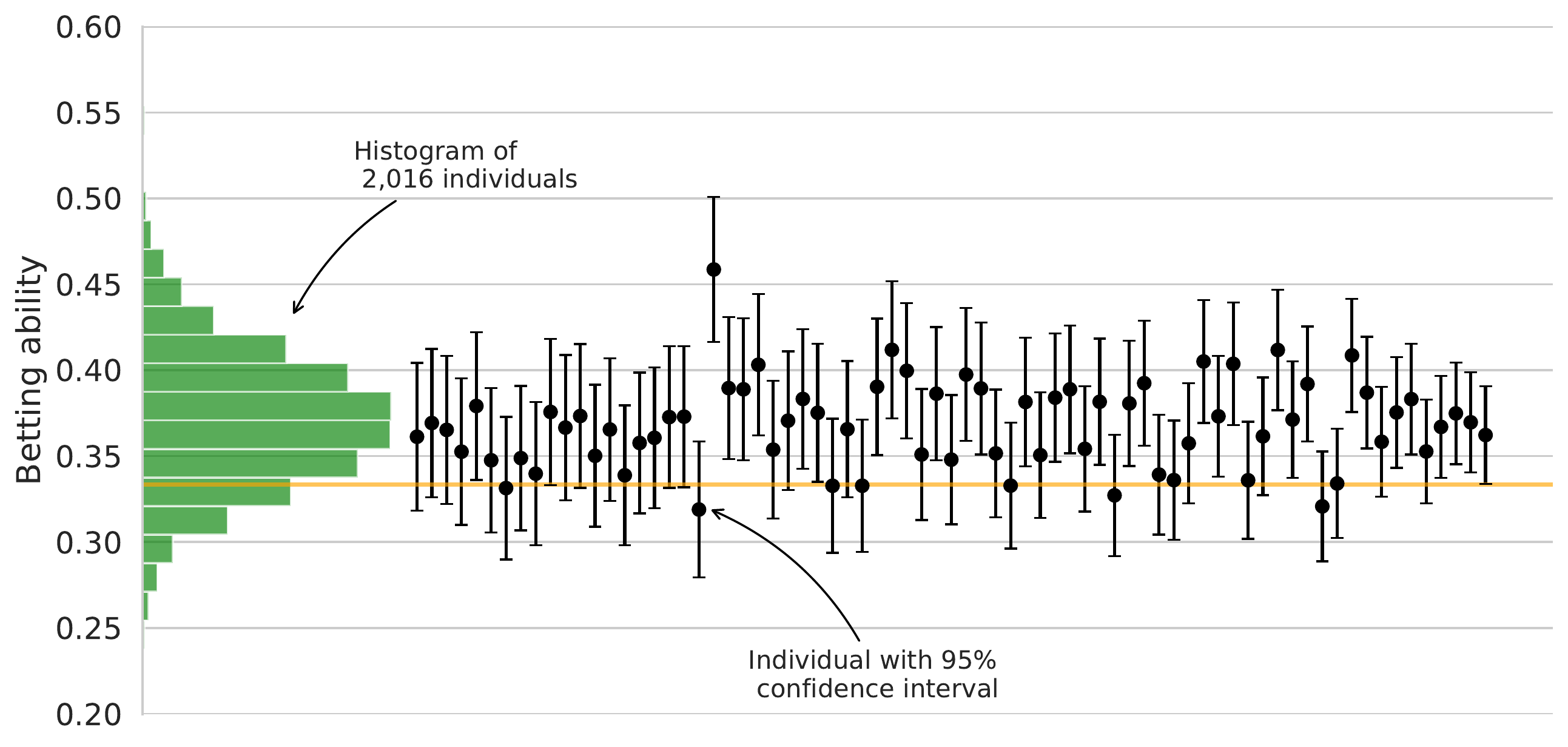}
 \begin{minipage}{1\textwidth}
    \footnotesize
    \emph{Notes:}  The histogram drawn on the vertical axis shows the distribution of betting ability (defined as the long-run proportion of correct predictions) for the 2\,016 gamblers for whom we observe bets on at least 100 matches. To the right of the histogram, we show the subsample of gamblers for whom we observe bets on at least 500 matches. For each of these gamblers, we plot their estimated ability with a black dot and the Clopper-Pearson 95 percent confidence interval as a line with whiskers. The fact that many of the confidence intervals do not intersect illustrates heterogeneity in ability. The horizontal line at $\frac{1}{3}$ shows the fraction of correct predictions that would result from random guessing.
      \end{minipage}

\end{figure}

To sharpen intuition, Figure~\ref{fig:ability} shows the expected ability for the subsample of gamblers with at least 500 predictions---for whom our ability estimates are most precise.  The figure orders individuals from left to right by the number of matches. As the number of matches increases, the confidence interval around the estimates becomes smaller. We show a histogram of the estimated ability for the larger group of 2\,016 gamblers for whom we observe predictions on at least 100 matches. We plot the expected ability from random guessing with a horizontal line. The expected ability from random guessing is $\frac{1}{3}$ as each match has three possible outcomes: home team wins, away team wins, or draw. The gamblers who have a confidence interval above this line, have a better betting ability than random guessing. In this sample, 47 percent do better than random guessing.

The figure highlights the heterogeneity in gambling ability. We can observe several gamblers whose confidence intervals do not intersect, which shows that some gamblers have higher ability than others. Unlike many other types of gambling, sports betting has a role for skill.


\section{Empirical Evidence of Learning}\label{section:learning}

To study how a gambler responds to feedback on his ability, we must separate the positive signal of a correct prediction from the income effect of a winning bet. SportPesa's midweek and weekend jackpots provide an ideal setting to separate these effects. The midweek jackpot awards prizes for 10 or more correct predictions out of 13 matches and the weekend jackpot awards prizes for 12 or more correct predictions out of 17 matches. Any variation in the number of correct predictions below 10 for the midweek jackpot and below 12 for the weekend jackpot has no income effect. We can isolate the effect of feedback by focusing on this range of the jackpot results. We note that there is likely an even stronger signal of ability when the gambler wins money---but we cannot use this variation because it is confounded by the direct effect of the money won.

The probability of winning a prize for the jackpots is very low, so focusing on the range of results below the cutoff for prizes still provides us with ample variation in signals of ability. If a gambler selected the favorite (the team most likely to win) for every match in the jackpot, he would have a less than one percent chance of winning a prize. In our dataset of bets, 0.68 percent of midweek jackpot bets and 0.94 percent of weekend jackpot bets won a prize.

\subsection{Gamblers Respond to Past Results}\label{section:first_stage}

We study gamblers' betting behavior in response to the share of correct predictions on the jackpot in the previous week. We use the following specification: %
\begin{align}
Betting_{it} =& \beta \ Jackpot_{i(t-1)} + \gamma \ Jackpot \ Tickets_{i(t-1)} + v_i + \epsilon_{it}
\end{align}
where $v_i$ are individual fixed effects.%
\footnote{Our main specification includes individual fixed effects to control for individual differences in betting ability and the propensity to gamble. Following recent work by \citet{imai2020} and others, we do not include time fixed effects as the use of individual and time fixed effects can obscure interpretation, except under conditions of linearly additive effects \citep{kropko2020,dechaisemartin2020}. For reference, we show results with both fixed effects in Appendix Table~\ref{table:pos_vs_neg_twoway}.}

We use two measures of betting behavior as dependent variables. First, we use an indicator for placing either a midweek or weekend jackpot bet. Second, we use the inverse hyperbolic sine (which is similar to a log transformation) of mobile money transfers to betting accounts. $Jackpot_{i(t-1)}$ measures the share of correct predictions on jackpot bets in the previous week. We are interested in the magnitude of $\beta$, the impact in the share of correct predictions in the previous week on the propensity to bet in the current week. We control for $Jackpot \ Tickets_{i(t-1)}$, which indicates the number of jackpot tickets purchased in the previous week.

\begin{table}[tb] \centering
  \caption{Response to previous week's betting results}
  \label{table:learning}
\scalebox{0.9}{
\begin{threeparttable}
\begin{tabular}{@{\extracolsep{5pt}}lcc}
\\[-1.8ex]\hline
\hline \\[-2.4ex]
 & \multicolumn{2}{c}{\textit{Dependent variable:}} \\
\cline{2-3}
\\[-3ex] & Placed jackpot bet & Betting expenditure \\
 & (1) & (2)\\
\hline \\[-3ex]
 Share correct in week $t-1$ & 0.148 & 0.424 \\
  & (0.017) & (0.095) \\
& [0.116, 0.180] & [0.237, 0.610] \\
\\[-3ex]
\hline \\[-3ex]
Individual fixed effects & Yes & Yes \\
Week fixed effects & No & No \\
Individuals & 15\,715 & 15\,715 \\
Weeks & 119 & 119 \\
Observations & 70\,390 & 70\,390 \\
\hline
\hline \\[-2.4ex]
\end{tabular}
\begin{tablenotes}
\item \small \emph{Notes:} All specifications control for the number of midweek jackpot and weekend jackpot tickets. The sample excludes individual-week observations where the individual won a prize on the jackpot in the previous week. Robust standard errors are shown in round brackets and the 95 \% confidence interval is shown in square brackets.
\end{tablenotes}
\end{threeparttable}
}
\end{table}

Results, shown in Table~\ref{table:learning}, indicate that gamblers react significantly to the previous week's betting results. A one standard deviation increase in the share of correct predictions in the previous week increases the probability that the gambler plays the jackpot in the current week by 1.78 percentage points and increases mobile money transfers to the betting account by 5.01 percent on average.


\FloatBarrier
\subsection{Symmetric Reaction to Positive and Negative Feedback}
\label{section:pos_vs_neg_feedback}

Many experiments find biased learning from feedback. People react to positive feedback and ignore or forget negative feedback---especially when feedback involves measures of intelligence or performance.%
\footnote{See, for example, \citet{eil2011good, sharot2011, mobius2014,godker2019, zimmerman2020,huffman2020} and \citet{chew2019motivated}.} %
In contrast, we find that gamblers react symmetrically to positive and negative feedback.

To study the difference between positive and negative feedback, we construct a categorical variable from the share of correct predictions on jackpot matches. We start by calculating the mean share of correct predictions for each individual. We then define three categories: (i) positive feedback as more than 10 percent above the mean, (ii) negative feedback as 10 or more percent below the mean, and (iii) between 10 percent above or below the mean as the base category. In Appendix~\ref{appendix:robustness} we show that results are not sensitive to the choice of a 10 percent threshold.%
\footnote{We restrict the analysis to gamblers for whom we observe at least one week in all three categories. Otherwise, the coefficients on the positive and negative feedback indicators would be estimated on different sets of individuals. We explain this choice in more detail in Appendix~\ref{appendix:biased_learning}.}

We test for biased learning using the following specification: %
\begin{equation}
\begin{aligned}
Bet \ on \ jackpot_{i(t+\tau)} =& \beta_p \ Positive \ Feedback_{it} + \beta_n \ Negative \ Feedback_{it} + \\
& \gamma \ Jackpot \ Tickets_{it} + v_i + \epsilon_{i(t+\tau)}
\end{aligned} %
\end{equation}
where $v_i$ are individual fixed effects and we control for the number of jackpot tickets purchased in a given week. We are interested in comparing the coefficients $\beta_p$ and $\beta_n$ to compare how gamblers respond to positive versus negative feedback of their betting ability. We investigate how betting results in week $t$ change the likelihood the gamblers places a jackpot bet in week $t+1$, $t+2$, $t+3$ and $t+4$.

\begin{table}[tbp] \centering
\caption{Response to positive and negative feedback}
\label{table:pos_vs_neg}
\scalebox{0.9}{
\begin{threeparttable}
\begin{tabular}{@{\extracolsep{5pt}}lcccc}
\hline
\hline \\[-1.8ex]
 & \multicolumn{4}{c}{Place jackpot bet in week:} \\
\cline{2-5}
\\[-1.8ex] & $t+1$ & $t+2$ &$t+3$ &$t+4$ \\
\hline \\[-1.8ex]
Positive feedback ($\beta_p$) & 0.028 & 0.011 & 0.014 & 0.008 \\
         & (0.005) & (0.006) & (0.006) & (0.006) \\
 Negative feedback ($\beta_n$) & -0.029 & -0.011 & -0.016 & -0.012 \\
         & (0.005) & (0.006) & (0.005) & (0.006) \\
  \hline \\[-1.8ex]
 P-value $|\beta_p| = |\beta_n|$ & 0.433 & 0.487 & 0.349 & 0.281 \\
 Individual Fixed Effects & Yes & Yes & Yes & Yes \\
 Week Fixed Effects & No & No & No & No \\
 Individuals  & 4\,399 & 4\,111 & 3\,900 & 3\,704 \\
 Weeks & 119 & 118 & 117 & 116 \\
 Observations & 48\,948 & 45\,626 & 43\,216 & 40\,959 \\
\hline
\hline \\[-2.4ex]
\end{tabular}
\begin{tablenotes}
\item \small \emph{Notes:} Dependent variable is an indicator for whether the individual places a jackpot bet in the weeks following a week of positive feedback or negative feedback on their gambling ability. Positive (or negative) feedback is defined as when the fraction of correct predictions made by the individual is more than 10\% higher (or lower) than their average rate of correct predictions.  All specifications control for the number of midweek jackpot and weekend jackpot tickets. The sample excludes individual-week observations where the individual won a prize on the jackpot in week $t$. We report robust standard errors in parenthesis below each estimate.
\end{tablenotes}
\end{threeparttable}
}
\end{table}

Table~\ref{table:pos_vs_neg} shows our estimates of the positive and negative feedback effects. Gamblers are more likely to bet following positive feedback and less likely to bet following negative feedback of their gambling ability. Remarkably, the magnitude of the coefficients are very similar in all specifications.%
\footnote{Since gamblers are less likely to bet following negative feedback, we may be concerned that the results are biased by changes in the composition of the sample over time. However, if this bias is present, our conclusion will be strengthened. Gamblers who do not respond to negative feedback are more likely to continue gambling. Thus, the sample is more likely to contain these gamblers.} %
The effect remains statistically significant for three weeks.

As this symmetry was unexpected, we surveyed 32 people in which we asked the question,
\textit{
``Each week a gambler bets on a number sports matches. How will the gambler respond to the past week's betting results?''} The possible responses are shown below, along with the number of respondents who selected that answer in parentheses.
\begin{itemize}
    \item \textit{The gambler will be more responsive to a positive result in the previous week than a negative result.} (18)
    \item \textit{The gambler will respond equally to positive and negative feedback.} (4)
    \item \textit{The gambler will be more responsive to a negative result in the previous week than a positive result.} (8)
    \item \textit{The gambler will not respond to the past week's betting results.} (2)
\end{itemize}
Only 4 respondents predicted our result while the majority predicted that gamblers would be more responsive to positive than to negative feedback.

Our model helps to highlight the implications of this result. The gambler forms a prior of his betting ability by watching $m_w$ matches and making $s_w$ successful predictions. Then, he bets on $m_b$ matches and makes $s_b$ successful bets. The gambler's expected betting ability is
\begin{align*}
    \mathrm{E}[\theta|s_b] &= \dfrac{s_w + s_b}{m_w+m_b} \ .
\end{align*}

Our empirical results study how gamblers respond to variation to $s_b$. Notice that the expected ability $\mathrm{E}[\theta|s_b]$ is linear in $s_b$. Variation in $s_b$ will change the gambler's expected ability by a constant factor, which is consistent with the similar magnitude in response to low values (negative feedback) and high values (positive feedback) of $s_b$. These results provide evidence against biased updating models where gamblers may discount or ``explain away" negative feedback \citep{gilovich1983biased}.

\section{How Do Gamblers Fund Gambling Expenditure?}\label{section:welfare}

Our final set of results explore the causal impact of gambling on the financial decisions of gamblers. Policymakers worry that gamblers may use credit to fund betting---a recipe for financial ruin. Anecdotes suggest that betting increases the demand for loans and causes bankruptcy, but we are not aware of causal evidence of the impact of increased betting expenditure on gamblers' other financial behaviors.%
\footnote{For instance, \citet{dahir2017} notes that "gambling addiction is on the rise in Kenya and leaving young people bankrupt and suicidal." An article by \citet{economist2018} warns that sports betting may be linked to high default rates on consumer loans: ``Anecdotal evidence is mounting of abuses---most notoriously of young Kenyans borrowing to splurge on online betting sites."}

\subsection{Identification and Estimation}

We are interested in estimating the causal effect of increased betting expenditure on the use of savings and credit that we observe in our data. Since betting expenditure, in general, is not random, we use an instrumental variables strategy. For an instrument to be valid in this context, it must be relevant (i.e., correlated with betting), and it must satisfy the exclusion restriction that it should be related to the use of savings and credit only through the endogenous measure of betting.

We use the share of correct jackpot predictions in week $t-1$ as an instrument for betting expenditure in week $t$. In Section~\ref{section:first_stage}, we demonstrated the relevance of this instrument by showing that an increase in the number of correct predictions on SportPesa's jackpot increased the propensity to bet in the following week. The second specification in Table~\ref{table:learning} shows the strong relationship between our instrument, the share of correct predictions in the previous week, and the endogenous variable, betting expenditure. The partial F-statistic for this specification is 19.87, well above the standard benchmark of 10 \citep{stock2005testing}.

Our exclusion restriction requires that the number of correct predictions on the jackpot is random conditional on the gambler's ability (approximated with an individual-specific fixed effect) and the number of jackpot tickets purchased (the endogenous regressor). In our context, it is difficult to imagine how outcomes may be impacted by the previous week's jackpot results other than through the current week's betting behavior.

One concern with the exclusion restriction is that winnings from prior weeks could directly impact outcomes in future weeks. However, as noted in Section~\ref{section:jackpots}, we only consider observations when the gambler's predictions fell below the threshold where the gambler wins money (10 in the midweek jackpot and 12 in the weekend jackpot). We exclude these observations from the sample to ensure our instrument is not driven by an income effect. Since the probability of winning money from a jackpot bet is small, the sample size reduces by less than one percent. 

Empirically, we estimate the impact of increased betting expenditure on measures of gambler $i$'s use of savings and credit at time $t$, denoted by $Y_{it}$, as:
\begin{align}
\label{eq:iv}
    Y_{it} =& \beta \arsinh(\widehat{Betting \ expenditure_{it}}) + v_i + \gamma \ jackpot \ tickets_{i(t-1)} + \epsilon_{it}
\end{align}
where $v_i$ are individual fixed effects and betting expenditure is instrumented by the share of correct jackpot predictions in the previous week. For all non-negative continuous outcomes, we use the inverse hyperbolic sine transformation, $\arsinh = \ln (x+\sqrt{x^{2}+1})$, and interpret  $\beta$ as the elasticity between betting expenditure and the outcome \citep{bellemare2020elasticities}.

We focus our analysis on a few specific margins of financial account use that we observe in our data:
\begin{itemize}
    \item \emph{Savings withdrawals:} The value of withdrawals from an individual's M-Shwari savings account during the week, in Kenyan Shillings (KSH), scaled with the inverse hyperbolic sine transformation. M-Shwari is the digital banking service offered by M-Pesa, Kenya's dominant mobile money service \citep{finaccess2019}.
    \item \emph{Savings deposits:} The value of deposits into the individual's M-Shwari account in a given week (KSH), scaled with the inverse hyperbolic sine transformation.
    \item \textit{Net savings deposits:} The value, in KSH, of all M-Shwari deposits minus the value of all withdrawals.
    \item \textit{Applied for a loan:} An indicator for whether the individual applied for a loan from one of several popular lending companies.%
    \footnote{The list of companies includes: M-Shwari, Tala, Branch, KCB, Equity Bank and Co-op Bank.}
    \item \emph{Loans received:} The value of loans received in a given week, defined as a payment from a loan company to the individual's mobile money account, scaled by the inverse hyperbolic sine transformation.
    \item \emph{Loan repayments:}  The value of loans repaid in a given week, defined as a payment from the individual to a loan company, scaled by the inverse hyperbolic sine transformation.
\end{itemize}

This is not the full set of savings and credit options available to gamblers---they could be transacting on accounts that are not mediated by their phone---but mobile money is the primary formal financial ecosystem used by most Kenyans, and one of the key drivers of financial inclusion in Kenya.%
\footnote{For instance, nationally representative survey evidence by \citet{finaccess2019} indicates that while 79\% of Kenyans have mobile money accounts, only 30\% have bank accounts.}

\subsection{Results}

Instrumental variables estimates of the impact of betting expenditures on the use of savings and credit are presented in Table~\ref{table:iv_results}. In columns 1 and 2, we find that increases in gambling expenditures cause gamblers to more actively use their savings accounts---both increasing the value of withdrawals and deposits to their accounts. Specifically, a one percent increase in gambling expenditure increases withdrawals from a savings account by 0.547 percent and increases top-ups into the savings account by 0.388 percent. The increase in withdrawals is of similar magnitude to the increase in deposits, and we cannot reject the null hypothesis that there is no effect on net savings accumulation (column 3).

Increases in gambling expenditures do not have a statistically significant impact on borrowing behavior. The estimates in columns 4-6 are imprecise, but we can reject with 95 confidence that a one percent increase in gambling expenditure would increase loan applications by more than 0.10 percentage points.

Taken together, the results in Table~\ref{table:iv_results} suggest that gamblers are not relying primarily on debt to fund their gambling activities. If anything, gamblers appear to be paying for gambling with the balance in their savings account---but without a clear negative effect on their net deposits.

It should be noted that our instrumental variables strategy identifies a local average treatment effect, and should thus be interpreted as the causal effect of relatively small increases in betting expenditures. Large shocks that dramatically alter a gambler's betting expenditures may have qualitatively different effects on financial behavior.

\begin{landscape}
\thispagestyle{empty}
\begin{table}[!htbp] \centering
  \caption{Instrumental variables estimates of the impact of increased betting expenditure}
  \label{table:iv_results}
\scalebox{0.9}{
\begin{threeparttable}
\begin{tabular}{@{\extracolsep{5pt}}lcccccc}
\\[-1.8ex]\hline
\hline \\[-1.8ex]
 & \multicolumn{6}{c}{\textit{Dependent variable:}} \\
\cline{2-7}
\\[-1.8ex] & (1) & (2) & (3) & (4) & (5) & (6)
\\& Savings & Savings  & Net savings &  Applied & Loans & Loan  \\[-1ex]
& withdrawal & deposit & deposit & for loan & received & repayments  \\
 \textit{Units} & \textit{Elasticity} & \textit{Elasticity} & \textit{KSH} & \textit{Indicator} & \textit{Elasticity} & \textit{Elasticity} \\
\hline \\[-1.8ex]
Betting expenditure & 0.547 & 0.388 & 124.358 & 0.025 & $-$0.204 & $-$0.442 \\
& (0.202) & (0.187) & (353.13) & (0.038) & (0.316) & (0.325) \\
& [0.151, 0.944] & [0.021, 0.754] & [$-$567.77, 816.49] & [$-$0.049, 0.099] & [$-$0.823, 0.416] & [$-$1.080, 0.196] \\
\hline \\[-1.8ex]
Individual fixed effects & Yes & Yes & Yes & Yes & Yes & Yes \\
Week fixed effects & No & No & No & No & No & No \\
Individuals & 15\,715 & 15\,715 & 15\,715 & 15\,715 & 15\,715 & 15\,715 \\
Weeks & 119 & 119 & 119 & 119 & 119 & 119 \\
Observations & 70\,390 & 70\,390 & 70\,390 & 70\,390 & 70\,390 & 70\,390 \\
\hline
\hline \\[-1.8ex]
\end{tabular}
\begin{tablenotes}
\item \small \emph{Notes:} Independent variable is the inverse hyperbolic sine of betting expenditures, instrumented with the past week's jackpot performance. The dependent variable in columns (1) and (2) are the inverse hyperbolic sine of withdrawals from and deposits to the M-Shwari savings account. The dependent variable in column (3) is the total deposits minus withdrawals, in Kenyan Shillings (KSH), with an approximate exchange rate of  100 KSH to \$1 USD. Dependent variables in columns (4)-(6) capture loan behavior on all loan providers who transact with mobile money, including M-Shwari, Branch and Tala. All specifications control for the number of midweek jackpot and weekend jackpot tickets purchased in the previous week. The sample excludes individual-week observations where the individual won a prize on the jackpot in the previous week. We report robust standard errors in round brackets and the 95 percent confidence intervals in square brackets.
\end{tablenotes}
\end{threeparttable}
}
\end{table}
\end{landscape}

\section{Conclusion}

Our analysis indicates that sports bettors differ in ability, react to past results, and react symmetrically to positive and negative feedback. We also provide carefully identified, although imprecise, estimates of the impact of increased betting expenditure on other types of financial activity. We do not find strong support for the hypothesis that sports betting systematically drives people into financial ruin.

Our results contradict a common intuition that gamblers continue betting without any regard for past performance, or that they react asymmetrically to wins and losses. Gamblers do learn from experience. However, this learning may require a large number of bets before the gambler has an accurate understanding of his or her ability.

\clearpage

\bibliographystyle{aernobold}
{\small
\bibliography{references}
}

\clearpage
\pagebreak
\appendix

\begin{center}
    \huge{\textbf{Online Appendix}}
\end{center}

\thispagestyle{empty}

\section{Descriptive Statistics}
\label{appendix:descriptive_stats}

Table~\ref{table:descriptive_stats} provides descriptive statistics of our dataset. Our sample consists of mostly young men. Among the subsample of gamblers, 77 percent are men and 75 percent are 35 years of age or younger. Gamblers spend 481.67 KSH (approximately 4.81 USD) and receive 364.40 KSH (approximately 3.64 USD) from gambling on average per week.  

The distribution of gambling activity is fat-tailed, with the top 20 largest spenders betting an average of KSH 84\,349.41 (approximately 843 USD) per week. Betting income has fat tails because the largest prizes are awarded for bets with extreme odds. Also, betting income only measures withdrawals from the betting account. Small wins may fund subsequent bets rather than being withdrawn from the betting account.

\section{Popularity of Sports Betting in Kenya}
\setcounter{figure}{0}
\renewcommand{\thefigure}{A\arabic{figure}}
\setcounter{table}{0}
\renewcommand{\thetable}{A\arabic{table}}

\begin{figure}[ht]
\centering
\includegraphics[width=\textwidth]{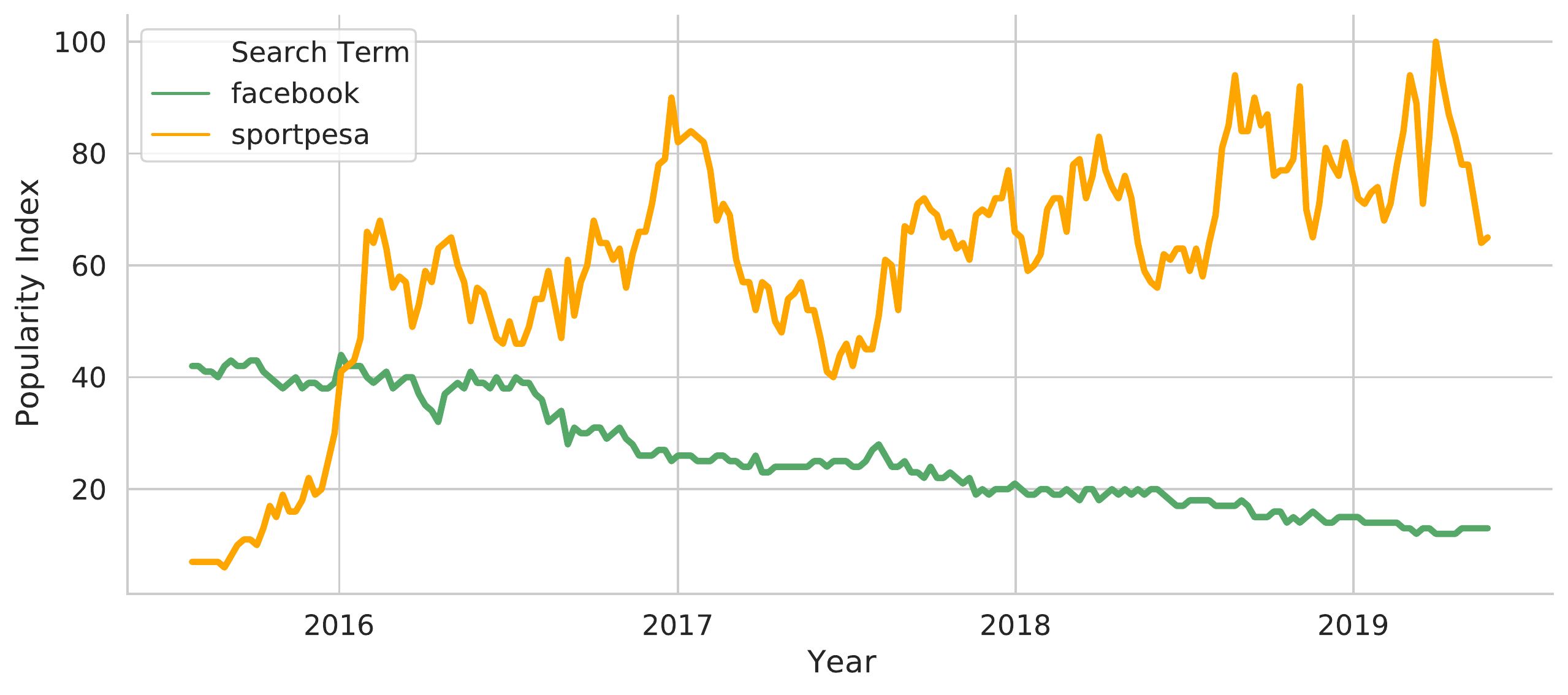}
\caption{Popularity of Google search queries in Kenya}
\label{fig:google_search}
\end{figure}

To emphasize the scale of sports betting, we use data on internet search queries on Google. In Figure \ref{fig:google_search} we plot the relative popularity of SportPesa against Facebook.%
\footnote{Visit \href{https://trends.google.com/trends/explore?date=all&geo=KE&q=sportpesa,facebook}{trends.google.com} to access a current version of the graph.} %
Facebook serves as a good benchmark as it is the most popular search query worldwide. Users of online services typically search for the name of the service rather than type out the web address so the index of search queries serves as a good proxy for the relative number of users. For example, a user may type in ``facebook" in the search bar rather than typing out ``www.facebook.com". Figure \ref{fig:google_search} shows a clear pattern. Since 2014, sports betting has grown rapidly in popularity. In 2018, SportPesa was the most popular search query in Kenya.

\FloatBarrier

\section{More Details on our Test for Biased Learning}
\label{appendix:biased_learning}

In Section \ref{section:pos_vs_neg_feedback} of the main paper, we test for biased learning using the following specification: %
\begin{align*}
Bet \ on \ jackpot_{i(t+\tau)} =& \beta_p \ Positive \ feedback_{it} + \beta_n \ Negative \ feedback_{it} + \\
& \gamma \ jackpot \ tickets_{it} \\
& v_i + \epsilon_{i(t+\tau)} 
\end{align*} %
The indicators for positive and negative feedback have three buckets. The share of correct predictions in week $t-1$ can be %
\begin{enumerate}
    \item positive (10 percent above the gambler's mean),
    \item negative (10 percent below the gambler's mean),
    \item or base (between 10 percent above and below the gambler's mean).
\end{enumerate}
We drop all gamblers who do not have at least one observation in each of the positive, negative, and base buckets. In this appendix, we explain why we need to restrict the sample in this way.

Suppose there are two types of gamblers: stubborn low skill $S$ and Bayesian high skill $B$. The $S$ gamblers get $L$ percent of predictions correct whereas $B$ gamblers get $H$ correct and $H > L$. 

We observe some bets and results for each type of gambler. Since $H > L$, the $B$ gambler is more likely to ``fill" the higher buckets of the categorical variable measuring the number of correct predictions in Week $t-1$. In contrast, the $S$ gambler is more likely to ``fill" the low buckets. 

If for a given gambler, one of the indicators is zero for all weeks we observe, this gambler does not contribute to the estimate of this coefficient. Therefore $S$ gamblers will contribute more to the estimation of the coefficients of the low result indicator and the $B$ gamblers will contribute more to the estimation of the high result indicators.

Assume $S$ gamblers do not consider past performance when betting and $B$ gamblers use rational Bayesian updating. This means that the higher bins will reflect Bayesian updating whereas the lower bins will reflect the stubborn betting. This will generate a biased updating result even though no single gambler is biased towards positive feedback.

\FloatBarrier

\section{Correlation in Correct Predictions Across Time}
\label{sec:dynamics}

The approach in Section \ref{section:ability} uses variation for the subsample of 73 gamblers for which we observe at least 500 match predictions. In this appendix, we use variation for the 6\,953 gamblers for which we observe jackpot bets in at least two consecutive weeks.

If all gamblers are identical and predict the outcome of matches with some fixed probability, there will be no correlation in the number of correct predictions across time. If a gambler who predicts a high number of matches correctly this week is more likely to predict a high number of matches correctly next week, this suggests heterogeneity in gambling ability. 

To test for correlation in the number of correct predictions across time, we use the following specification:
\begin{align*}
Correct \ predictions_{it} =& \sum_{l=1}^L\beta_l \ Correct \ predictions_{i(t-l)} \\
& + \theta_t + \sum_{l=1}^L \gamma_l \ jackpot \ tickets_{i(t-l)} +  \epsilon_{it} 
\end{align*}
The model checks if the number of correct predictions in week $t$ by gambler $i$ is positively correlated with the number of correct predictions in previous weeks up to a lag of $L$ weeks. We include week fixed effects, $\theta_t$, so the estimates use only within week variation to compare gamblers. We also control for the number of midweek and the number of weekend jackpot tickets purchased.

We report the results in Table \ref{table:het_ability}. In all specifications, the number of correct predictions on jackpot matches is positively correlated with the number of correct predictions in previous weeks. If all gamblers had identical chances of success, this correlation would not be present. The positive correlation can only be explained by heterogeneity in the chance of success between gamblers, which we interpret in our model as heterogeneity in gambling ability.

\section{Robustness Checks}
\label{appendix:robustness}

\textbf{Cutoffs for the low and high categories}

In Section \ref{section:pos_vs_neg_feedback}, we use 10 percent above and below the mean number of correct predictions as the cutoff to form the positive feedback, negative feedback, and base categories. Here, we test the robustness of our results to cutoffs of 5 percent and 15 percent. Note that, as described in Appendix~\ref{appendix:biased_learning}, each gambler must have at least one observation in each of the three categories. By adjusting the cutoff for the three categories, the number of gamblers used in the estimation changes slightly with the cutoff threshold. 

Results, presented in Tables~\ref{table:pos_vs_neg_05} and~\ref{table:pos_vs_neg_15} indicate that gamblers react to both positive and negative feedback. While the absolute value of the coefficient is often larger for negative feedback than for positive feedback, in most cases the difference in absolute magnitude is not statistically significant. \\

\noindent \textbf{Including time fixed effects}

We use individual fixed effects to control for individual differences in betting ability and other time-invariant characteristics. We choose not to use time fixed effects in addition to the individual effects. For reference, Table \ref{table:pos_vs_neg_twoway} shows the results with the inclusion of time fixed effects.
\\

\begin{table}[htbp] \centering
\caption{Descriptive statistics} 
\label{table:descriptive_stats} 
\scalebox{0.9}{
\begin{threeparttable} 
\begin{tabular}{@{\extracolsep{5pt}}lccccccc} 
\\[-1.8ex]\hline 
\hline \\[-1.8ex]
 & Gamblers & Non-Gamblers & Full sample \\
\hline \\[-1.8ex]
Number of individuals & 58\,215 & 35\,350 & 93\,565 \\ 
 Male & 77.0 \%  & 52.0 \%  & 67.6 \%  \\
 Age & 30.66 & 31.74 & 31.06 \\ 
  &  [25, 34] &  [25, 36] &  [25, 35] \\ 
 Betting expenditure (weekly) & 481.67 & 0.00 & 299.69 \\ 
  &  [18, 306] &  [0, 0] &  [0, 132] \\
 Betting income (weekly) & 364.40 & 0.00 & 226.73 \\ 
  &  [0, 177] &  [0, 0] &  [0, 46] \\
\hline 
\hline 
\end{tabular}
\begin{tablenotes} 
\item \small \emph{Notes:} Mean values reported, with 25th and 75th percentiles in brackets. We define gamblers as individuals with at least one mobile money transfer to or from a betting company. We exclude users with less than three weeks of mobile money transactions from the sample. Monetary amounts are measured in Kenyan Shillings (KSH). 
\end{tablenotes}
\end{threeparttable}
}
\end{table}

\begin{table}[!htbp] \centering 
  \caption{Testing for heterogeneity in betting ability} 
  \label{table:het_ability}
 \scalebox{0.9}{
 \begin{threeparttable}
\begin{tabular}{@{\extracolsep{5pt}}lcccc} 
\\[-1.8ex]\hline 
\hline \\[-1.8ex] 
 & \multicolumn{4}{c}{\textit{Dependent variable:}} \\ 
\cline{2-5} 
\\[-1.8ex] & \multicolumn{4}{c}{Share of correct predictions in week $t$} \\ 
\\[-1.8ex] & (1) & (2) & (3) & (4)\\ 
\hline \\[-1.8ex] 
 Share of correct & & & & \\
 predictions in week: & & & & \\
 $  t-1$ & 0.077 & 0.075 & 0.075 & 0.068 \\ 
  & (0.005) & (0.007) & (0.008) & (0.009) \\ 
 $  t-2$ &  & 0.064 & 0.062 & 0.056 \\ 
  &  & (0.007) & (0.008) & (0.009) \\ 
 $  t-3$ &  &  & 0.067 & 0.074 \\ 
  &  &  & (0.008) & (0.009) \\ 
 $  t-4$ &  &  &  & 0.059 \\ 
  &  &  &  & (0.009) \\ 
\hline \\[-1.8ex] 
Individual fixed effects & No & No & No & No \\ 
Week fixed effects & Yes & Yes & Yes & Yes \\ 
Individuals & 6\,953 & 4\,008 & 2\,589 & 1\,825 \\ 
Weeks & 119 & 118 & 117 & 116 \\ 
Observations & 35\,642 & 22\,370 & 15\,749 & 11\,854 \\  
\hline 
\hline \\[-1.8ex]
\end{tabular}
\begin{tablenotes}
\item \small \emph{Notes:} All specifications control for the number of midweek jackpot and weekend jackpot tickets. The sample excludes individual-week observations where
the individual won a prize on the jackpot in week $t$. Robust standard errors in parenthesis.
\end{tablenotes}
\end{threeparttable}
}
\end{table}

\begin{table}[tbp] \centering
\caption{Response to positive and negative feedback with a 5\% cutoff} 
\label{table:pos_vs_neg_05} 
\small 
\begin{threeparttable} 
\begin{tabular}{@{\extracolsep{5pt}}lcccc} 
\\[-1.8ex]\hline 
\hline \\[-1.8ex]
 & \multicolumn{4}{c}{Place jackpot bet in week:} \\ 
\cline{2-5} 
\\[-1.8ex] & $t+1$ & $t+2$ &$t+3$ &$t+4$ \\  
\hline \\[-1.8ex] 
Positive feedback ($\beta_p$) & 0.015 & 0.013 & 0.011 & 0.004 \\ 
         & (0.005) & (0.005) & (0.005) & (0.005) \\ 
 Negative feedback ($\beta_n$)& -0.026 & -0.011 & -0.010 & -0.009 \\ 
         & (0.005) & (0.005) & (0.005) & (0.005) \\ 
   \hline \\[-1.8ex]
  P-value $|\beta_p| = |\beta_n|$  & 0.0169 & 0.3959 & 0.413 & 0.200 \\ 
 Individual Fixed Effects & Yes & Yes & Yes & Yes \\ 
 Week Fixed Effects & No & No & No & No \\ 
 Individuals  & 5\,309 & 4\,953 & 4\,678 & 4\,454 \\ 
 Weeks & 119 & 118 & 117 & 116 \\ 
 Observations & 53\,601 & 49\,904 & 47\,196 & 44\,875 \\ 
\hline 
\hline \\[-1.8ex]
\end{tabular}
\begin{tablenotes} 
\item \footnotesize \emph{Notes:} Dependent variable is an indicator for whether the individual places a jackpot bet in the weeks following a week of positive feedback or negative feedback on their gambling ability. Positive (or negative) feedback is defined as when the fraction of correct predictions made by the individual is more than 10\% higher (or lower) than their average rate of correct predictions.  All specifications control for the number of midweek jackpot and weekend jackpot tickets. The sample excludes individual-week observations where the individual won a prize on the jackpot in week $t$. We report robust standard errors in parenthesis below each estimate.
\end{tablenotes}
\end{threeparttable} 
\end{table}

\begin{table}[tbp] \centering
\caption{Response to positive and negative feedback with a 15\% cutoff} 
\label{table:pos_vs_neg_15} 
\small 
\begin{threeparttable} 
\begin{tabular}{@{\extracolsep{5pt}}lcccc} 
\\[-1.8ex]\hline 
\hline \\[-1.8ex]
 & \multicolumn{4}{c}{Place jackpot bet in week:} \\ 
\cline{2-5} 
\\[-1.8ex] & $t+1$ & $t+2$ &$t+3$ &$t+4$ \\  
\hline \\[-1.8ex] 
Positive feedback ($\beta_p$) & 0.020 & 0.007 & 0.010 & 0.003 \\ 
         & (0.008) & (0.009) & (0.008) & (0.009) \\ 
 Negative feedback ($\beta_n$) & -0.033 & -0.015 & -0.013 & -0.017 \\ 
         & (0.008) & (0.008) & (0.008) & (0.008) \\ 
   \hline \\[-1.8ex]
  P-value $|\beta_p| = |\beta_n|$ & 0.117 & 0.239 & 0.380 & 0.101 \\ 
 Individual Fixed Effects & Yes & Yes & Yes & Yes \\ 
 Week Fixed Effects & No & No & No & No \\ 
 Individuals  & 2\,409 & 2\,233 & 2\,118 & 2\,024 \\ 
 Weeks & 119 & 118 & 117 & 116 \\ 
 Observations & 34\,027 & 31\,388 & 29\,646 & 28\,265 \\ 
\hline 
\hline \\[-1.8ex]
\end{tabular}
\begin{tablenotes} 
\item \footnotesize \emph{Notes:} Dependent variable is an indicator for whether the individual places a jackpot bet in the weeks following a week of positive feedback or negative feedback on their gambling ability. Positive (or negative) feedback is defined as when the fraction of correct predictions made by the individual is more than 10\% higher (or lower) than their average rate of correct predictions.  All specifications control for the number of midweek jackpot and weekend jackpot tickets. The sample excludes individual-week observations where the individual won a prize on the jackpot in the previous week. We report robust standard errors in parenthesis below each estimate.
\end{tablenotes}
\end{threeparttable} 
\end{table}

\begin{table}[tbp] \centering
\caption{Response to positive and negative feedback with time fixed effects} 
\label{table:pos_vs_neg_twoway} 
\small 
\begin{threeparttable} 
\begin{tabular}{@{\extracolsep{5pt}}lcccc} 
\\[-1.8ex]\hline 
\hline \\[-1.8ex]
 & \multicolumn{4}{c}{Place jackpot bet in week:} \\ 
\cline{2-5} 
\\[-1.8ex] & $t+1$ & $t+2$ &$t+3$ &$t+4$ \\  
\hline \\[-1.8ex] 
Positive feedback ($\beta_p$) & 0.035 & 0.016 & 0.015 & 0.009 \\ 
         & (0.006) & (0.006) & (0.006) & (0.006) \\ 
 Negative feedback ($\beta_n$) & -0.030 & -0.014 & -0.017 & -0.011 \\ 
         & (0.006) & (0.006) & (0.006) & (0.006) \\ 
   & & & & \\ 
  \hline \\[-1.8ex] 
 P-value $|\beta_p| = |\beta_n|$ & 0.277 & 0.402 & 0.381 & 0.391 \\
 Individual Fixed Effects & Yes & Yes & Yes & Yes \\ 
 Week Fixed Effects & Yes & Yes & Yes & Yes \\ 
 Individuals  & 4\,399 & 4\,111 & 3\,900 & 3\,704 \\ 
 Weeks & 119 & 118 & 117 & 116 \\ 
 Observations & 48\,948 & 45\,626 & 43\,216 & 40\,959 \\ 
\hline 
\hline \\[-1.8ex]
\end{tabular}
\begin{tablenotes} 
\item \footnotesize \emph{Notes:} Dependent variable is an indicator for whether the individual places a jackpot bet in the weeks following a week of positive feedback or negative feedback on their gambling ability. Positive (or negative) feedback is defined as when the fraction of correct predictions made by the individual is more than 10\% higher (or lower) than their average rate of correct predictions.  All specifications control for the number of midweek jackpot and weekend jackpot tickets. The sample excludes individual-week observations where the individual won a prize on the jackpot in the previous week. We report robust standard errors in parenthesis below each estimate.
\end{tablenotes}
\end{threeparttable} 
\end{table}

\end{document}